


\magnification=1200
\input harvmac.tex
\hfuzz=10pt


\def\a{\alpha}  
\def\d{\delta}  \def\l{\lambda}
\def\f{\phi^{\bar n}_{\bar j}}
\def\D{\Delta} \def\ve{\varepsilon}

\def\mat{{\rm Mat}} 	
\def\bbc{C\kern-6.5pt I}   \def\bbz{Z\!\!\!Z}
\def\sL{{$sl_q(3)$}} 	\def\SL{{$SL_q(3)$}}
 	\def\SU{{$SU_q(3)$}}

\def\z{z_{1}^{n_1} z_{2}^{n_2} z_{3}^{n_3}}
\def\t{t^{\bar n}}


\baselineskip=12pt
\line{\hfill SISSA 57/94/FM}
\line{\hfill hep-th 9405131}
\vskip 1.5cm
\baselineskip=16pt plus 2pt minus 1pt

\centerline{{\bf
LEFT REGULAR REPRESENTATION OF $sl_q(3)$~:}}
\vskip .5cm
\centerline{{\bf
REDUCTION AND INTERTWINERS}}

\baselineskip=12pt

\vskip 1.5cm

\centerline{{\bf Ludwik D\c abrowski} ~and ~{\bf Preeti Parashar}}

\vskip 0.8cm

\centerline{ SISSA, Strada Costiera 11, 33014 Trieste, Italy}
\bigskip

\baselineskip=16pt plus 2pt minus 1pt
\parskip=7pt plus 1pt
\vskip 2cm

\centerline{\bf Abstract}
\smallskip\midinsert\narrower\narrower\noindent
Reduction of the left regular representation of quantum algebra $sl_q(3)$
is studied and ~$q$-difference intertwining operators are constructed.
The irreducible representations correspond to the spaces of local sections of
certain line bundles over the q-flag manifold.
\endinsert

\vfill\eject
\baselineskip=16pt plus 2pt minus 1pt
\parskip=7pt plus 1pt
\newsec{Introduction.}
Recently, following the canonical procedure for $q=1$
\ref\Dob{V.K. Dobrev, Rep. Math. Phys. {\bf 25} (1988) 159.},
the left regular representation, its reduction to infinite family
of (reducible and irreducible) representations and the ~$q$-difference
intertwining operators were studied on the examples of Lorentz quantum algebra
\ref\DDF{L. D\c abrowski, V.K. Dobrev, R. Floreanini, J. Math. Phys.
{\bf 35} (1994) 971.},
of $sl_q(2)$ and its contraction to $e_q(2)~$
\ref\DDF{L. D\c abrowski, J. Sobczyk, Lett. Math. Phys., in print}.
In this paper we investigate the case of $sl_q(3)$,
a quantization of the simple Lie algebra of rank two.
This Hopf algebra, besides being computationally more involved, presents some
new features which are interesting enough in order to generalize to the
case of $sl_q(n)$ and next to q-deformations of all semisimple Lie algebras.
This will help to understand better the important relation between
representation theory and geometry of quantum groups.

Denote by $G_q$ the quantum group ${\rm Fun}_q(G)$
and by $g_q$ its dual  $U_q(g)$, a quantization of the Lie algebra $g$ of $G$.
Recall that the left regular representation
$L : g_q \times G_q \to G_q $,
$(a, \varphi ) \to {\cal L}(a)\varphi $,
and the right regular representation
$R : g_q \times G_q  \to G_q $,
$(a, \varphi ) \to {\cal R}(a)\varphi $,
of $g_q$ are defined on $G_q$ by, respectively
\eqn\lerere{ \bigl( {\cal L}(a)\varphi \bigr) (b)=\varphi (S(a)b)}
\eqn\rirere{ \bigl( {\cal R}(a)\varphi \bigr) (b)=\varphi (ba),}
where $a, b\in g_q$, $\varphi\in G_q$ and $S$ is the antipode.

We shall reduce the left regular representation on the
eigenspaces of the right regular representation of the Borel generators.
Namely, we shall impose the condition that the Cartan
generators act as a multiplication by fixed numbers (to be specified further)
and that the other Borel generators vanish.
Then some of the remaining generators of $g$ yield (via the right regular
representation)
interesting intertwining q-difference operators, the kernels of which provide
a further reduction.
This method yields a natural q-deformation of several equations of
mathematical physics, which possess a symmetry group and which can be viewed as
intertwiner between some representations of the Lie algebra of the symmetry.
It may be helpful also in the inverse problem of finding the symmetry behind
a given q-difference equation.
\newsec{Preliminaries on ~\sL~ and ~\SL~.}
The generators of ~\SL~ (quantization of the algebra of complex functions
on ~\SL~) are the unit $1$ and $t_{ij}$, with $i, j = 1, 2, 3$;
arranged as a $3\times 3$ matrix $T$, satisfying the commutation relations
\eqn\comrel{
R~ T_1 T_2 = T_2 T_1 ~R
 \qquad (T_1=T\otimes I ,\ T_2=I\otimes T)\ ,}
where
\eqn\rmat{
q^{{1 \over 3}} R =
q\sum_i^3 e_{ii}\otimes e_{ii}
+ \sum_{i\ne j}^3  e_{ii}\otimes e_{jj}
+ \l \sum_{i>j}^3 e_{i,j}\otimes e_{j,i} \ ,}
with $\l = q-q^{-1}$. More explicitly, we have:
\eqn\coma{\eqalign{&
t_{11} t_{12} = q t_{12} t_{11},~ t_{11} t_{13} = q t_{13} t_{11},~
t_{11} t_{21} = q t_{21} t_{11},
\cr &
t_{11} t_{31} = q t_{31} t_{11},~
[t_{11},t_{23}] = \l t_{13} t_{21},~ [t_{11},t_{32}] = \l t_{12} t_{32},
\cr } }
\eqn\comb{\eqalign{&
t_{22} t_{12} = q^{-1} t_{12} t_{22},~ t_{22} t_{13} = t_{13} t_{22},~
t_{22} t_{23} = q t_{23} t_{22},
\cr &
t_{22} t_{21} = q^{-1} t_{21} t_{22},~
t_{22} t_{31} = t_{31} t_{22},~ t_{22} t_{32} = q t_{32} t_{22},
\cr } }
\eqn\comc{\eqalign{&
[t_{33},t_{12}] = -\l t_{13} t_{32},~ t_{33} t_{13} = q^{-1} t_{13} t_{33},~
t_{33} t_{23} = q^{-1} t_{23} t_{33},
\cr &
[t_{33},t_{21}] = -\l t_{23}t_{31},~
t_{33} t_{31} = q^{-1} t_{31} t_{33},~ t_{33} t_{32} = q^{-1} t_{32} t_{33},
\cr } }
\eqn\com{\eqalign{&
[t_{11},t_{22}] = \l t_{12}t_{21},~ [t_{11},t_{33}] = \l t_{13}t_{31},~
[t_{22},t_{33}] = \l t_{23}t_{32},
\cr } }
\eqn\comab{\eqalign{&
t_{12} t_{13} = q t_{13} t_{12},~ t_{12} t_{21} = t_{12} t_{21},~
[t_{12},t_{23}] = \l t_{13}t_{22},~
t_{12} t_{31} = t_{31} t_{12},~ t_{12} t_{32} = q t_{32} t_{12},
\cr } }
\eqn\comac{\eqalign{&
t_{13} t_{21} = t_{21} t_{13},~ t_{13} t_{23} = q t_{23} t_{13},~
t_{13}t_{31} = t_{31} t_{13},~ t_{13}t_{32} = t_{32}t_{13},
\cr } }
\eqn\comba{\eqalign{&
t_{21} t_{23} = q t_{23} t_{21},~ t_{21} t_{31} = q t_{31} t_{21},~
[t_{21},t_{32}] = \l t_{31}t_{22},
\cr } }
\eqn\combc{\eqalign{&
t_{23}t_{31} = t_{31} t_{23},~ t_{23}t_{32} = t_{32} t_{23},
\cr } }
\eqn\comca{\eqalign{&
t_{31} t_{32} = q t_{32} t_{31}.
\cr } }
An additional relation is
\eqn\det{
Det_q =
t_{11} (t_{22}t_{33} - q t_{23}t_{32})
- q t_{21} (t_{12}t_{33} - q t_{13}t_{32})
+ q^2 t_{31} (t_{12}t_{23} - q t_{13}t_{22})
= 1}
which can be written also in equivalent form as
\eqn\dete{
Det_q =
t_{11} (t_{22}t_{33} - q t_{23}t_{32})
- q t_{12} (t_{21}t_{33} - q t_{23}t_{31})
+ q^2 t_{13} (t_{21}t_{32} - q t_{22}t_{31})
= 1}
Considered as a Hopf algebra, \SL \  has the following comultiplication $\D$,
counit $\ve$, and antipode $S$ given on the generators by:
\eqn\tcop{\D t_{ij} = t_{ik}\otimes t_{kj}}
\eqn\tcou{\ve t_{ij} = \d _{ij}}
\eqn\tant{S t_{ij} = (T^{-1})_{ij}.}
For $\bar q = q,$ with the $*$-conjugation (complex anti-linear algebra
anti-involution and co-algebra involution)
$t_{ij}^* = S t_{ji}$, ~\SL \ becomes a Hopf $*$-algebra denoted ~\SU\ .

As generators  of the algebra \sL , dual to \SL , we shall use the unit $1$
and the functionals which have been introduced in
\ref\FRT{L. D. Faddeev, N. Yu. Reshetikhin and L. A. Takhtajan,
{Algebraic Analysis} {\bf 1} (1988) 129.}:
$l^{+}_{ij}$ with $i<j$; $l^{-}_{ij}$ with $i>j$; $l^{+}_{ii}$ and
$l^{-}_{ii} $, where $i, j = 1, 2, 3$.
When arranged in upper-~ and lower-triangular matrices $L^{\pm }$,
respectively, they are defined by the duality conditions
\eqn\dual{
(L^{\pm}, ~ T_1 \dots T_m ) =
R_1^{\pm } \dots R_m^{\pm }\ , \ \  {\rm for}~ m = 1, 2,\dots ;}
where for $1 \leq \ell \leq m$,~
$T_{\ell}^{\pm }$ act in the ${\ell}^{th}$ factor;
$R_{\ell}^{\pm }$ act in the $(0, \ell )^{th}$ factor of
$(\bbc ~^3 )^{\otimes (m+1)}$;
$R^{+} = P R P\ $, $R^{-} = R^{-1}\ $
and $P\in \mat (\bbc ~^3 \otimes \bbc ~^3 )$ is the permutation matrix.

The commutation relations are
\eqn\lcomrel{
R^{+} L_1^{\pm} L_2^{\pm }=L_2^{\pm }L_1^{\pm }R^{+}
\ ,\quad
R^{+} L_1^{+} L_2^{-}=L_2^{-}L_1^{+} R^{+} \ .}
Additional relations are:
\eqn\relpm{
l^{+}_{ii} l^{-}_{ii} = 1, \ i=1,2,3\ ;}
\eqn\detl{
l^{+}_{11} l^{+}_{22} l^{+}_{33} =
l^{-}_{11} l^{-}_{22} l^{-}_{33} = 1\ .}
The Hopf algebra structure is then given by
\eqn\lcop{\D l^{\pm}_{ij} = l^{\pm}_{ik} \otimes l^{\pm}_{kj}}
\eqn\lcou{\ve l^{\pm}_{ij} = \d _{ij}}
and the antipode which is defined as for $T$ with the replacement
of $q$ by $q^{-1}$.
The $l_{ij}^{\pm}$ can be expressed in terms of more popular generators
$q^{H_i}, X^{\pm}_i$, $i= 1, 2$ via:
\eqn\pop{ \eqalign{
q^{H_1} = l_{11}^+ (l_{22}^+)^{-1}, ~~&
q^{H_2} = l_{11}^+ (l_{22}^+)^{2}, 			\cr
X_1^+ = \l^{-1} q^{-1/6} l_{12}^+ (l_{11}^+)^{-1/2}(l_{22}^+)^{-1/2}, ~~&
X_2^+ = \l^{-1} q^{-1/6} l_{23}^+ (l_{11}^+)^{1/2}, \cr
X_1^- = -\l^{-1} q^{1/6} l_{21}^- (l_{11}^+)^{1/2}(l_{22}^+)^{1/2}, ~~&
X_2^- = -\l^{-1} q^{1/6} l_{32}^- (l_{11}^+)^{-1/2}. \cr} }
They satisfy the following commutation relations:
\eqn\pcom{ \eqalign{&
[H_i, H_j] = 0,~[H_i, X^{\pm}_j] = \pm (\alpha_i, \alpha_j)X^\pm_j,\cr
&[X^{+}_{i}, X^{-}_{j}] = \delta_{ij} [H_i]~,~i, j = 1, 2;\cr
&\sum^m_{k=0} (-1)^k \pmatrix{m\cr k}_q q^{k(k-m)(\alpha_i,\alpha_i)/2}
(X^{\pm}_{i})^k X^{\pm}_{j} (X^{\pm}_{i})^{m-k} = 0 for i\ne j\ ,\cr} }
where $[H_i] = {(q^{H_i}\! -\! q^{-H_i})\over (q\! -\! q^{-1})}\ $,
$m = 1 - A_{ij}$ and
$\pmatrix{\! m\cr k\! }_q = {(q^m - 1)(q^{m-1} - 1) \cdots
(q^{m-k+1} -1)\over {(q^k -1)(q^{k-1} -1) \cdots (q-1)}}.$

The Hopf algebra structure is then given by
\eqn\pcop{\D (H_i)~=~H_i\otimes 1 + 1\otimes H_i,\
\D (X_i^{\pm})~=~X_i^{\pm}\otimes q^{-{H_i/2}}~+~q^{H_i/2}\otimes X_i^{\pm}}
\eqn\pcou{\ve (X_i^{\pm~})=~0,\ \ve (H_i)~=0~}
\eqn\pant{S(X_i^{\pm}) = q^{-\rho } X_i^{\pm} q^{\rho },\  S(H_i)=-H_i }
with $\rho = {\sum H_{\a }}/2 $, where ${\a}$ belongs to the set of
positive roots.
\newsec{The left regular representation of ~\sL~ on ~\SL~.}
It can be easily seen from the definition \lerere~ that the left regular
representation of $l^{\pm}_{ij}$ amounts to a multiplication of the
generator matrix $T$ from the left by, respectively, the numerical matrices~:
\eqn\lreg{
\eqalign{&
(Sl^{+}_{11}, T) = q^{1\over 3} \!\pmatrix{{1\over q}&0&0\cr 0&1&0\cr 0&0&1}\!
,
(Sl^{+}_{12}, T) = -q^{1\over 3} \l \!\pmatrix{0&0&0\cr 1&0&0\cr 0&0&0}\! ,
(Sl^{+}_{13}, T) = -q^{1\over 3} \l \!\pmatrix{0&0&0\cr 0&0&0\cr 1&0&0}\! ,\cr
& (Sl^{-}_{21}, T) = q^{-{1\over 3}} \l \!\pmatrix{0&1&0\cr 0&0&0\cr 0&0&0}\! ,
(Sl^{+}_{22}, T) = q^{1\over 3} \!\pmatrix{1&0&0\cr 0&{1\over q}&0\cr 0&0&1}\!
,
(Sl^{+}_{23}, T) = -q^{1\over 3} \l \!\pmatrix{0&0&0\cr 0&0&0\cr 0&1&0}\! , \cr
& (Sl^-_{31}, T) = q^{-{1\over 3}} \l \!\pmatrix{0&0&1\cr 0&0&0\cr 0&0&0}\! ,
(Sl^-_{32}, T) = q^{-{1\over 3}} \l \!\pmatrix{0&0&0\cr 0&0&1\cr 0&0&0}\! ,
(Sl^{+}_{33}, T) = q^{1\over 3} \!\pmatrix{1&0&0\cr 0&1&0\cr 0&0&{1\over q}}
\! .\cr}}
Similarly, the right regular representation amounts to a multiplication of
$T$ from the right by, respectively,
\eqn\rreg{
\eqalign{&
(T, l^{+}_{11}) = q^{-{1\over 3}} \!\pmatrix{q&0&0\cr 0&1&0\cr 0&0&1}\! ,
(T, l^{+}_{12}) = q^{-{1\over 3}} \l \!\pmatrix{0&0&0\cr 1&0&0\cr 0&0&0}\! ,
(T, l^{+}_{13}) = q^{-{1\over 3}} \l \!\pmatrix{0&0&0\cr 0&0&0\cr 1&0&0}\! ,\cr
& (T, l^{-}_{21}) = -q^{1\over 3} \l \!\pmatrix{0&1&0\cr 0&0&0\cr 0&0&0}\! ,
(T, l^{+}_{22}) = q^{-{1\over 3}} \!\pmatrix{1&0&0\cr 0&q&0\cr 0&0&1}\! ,
(T, l^{+}_{23}) = q^{-{1\over 3}} \l \!\pmatrix{0&0&0\cr 0&0&0\cr 0&1&0}\! ,\cr
& (T, l^{-}_{31}) = -q^{1\over 3} \l \!\pmatrix{0&0&1\cr 0&0&0\cr 0&0&0}\! ,
(T, l^{-}_{32}) = -q^{1\over 3} \l \!\pmatrix{0&0&0\cr 0&0&1\cr 0&0&0}\! ,
(T, l^{+}_{33}) = q^{-{1\over 3}} \!\pmatrix{1&0&0\cr 0&1&0\cr 0&0&q}\! .\cr}}
As an (overcomplete) basis of \SL \ we can take the following ordered
monomials:
\eqn\basis{
t^{\bar n}
=: t_{11}^{n_{11}}~ t_{12}^{n_{12}} \dots t_{33}^{n_{33}} 	\ ,}
where $\bar n =: (n_{11}, n_{12}, \dots , n_{33})$ with
$n_{ij}\in \bbz_+ $
and we use (unless otherwise stated) the usual ordering, i.e. first
according to the row index and, then, the column index.

In the sequel we shall need an {\sl independent} basis for ~\SL~
which we believe must be known, but we have not been able to find it
in the literature for $SL_q(n)$ with $n\geq 3$.
In order to obtain such an independent basis we have to restrict the elements
$t^{\bar n}$ by the $det_q =1$ condition \rmat~. We proceed as follows.

Assume for a moment another ordering according to which the maximal power
$(t_{11} t_{22} t_{33})^p$, where $p = {\rm min}\{n_{11}, n_{22}, n_{33}\}$,
of product of diagonal generators $t_{11}$, $t_{22}$ and $t_{33}$
come first, then the remaining powers of (at most two of) $t_{ii}$ and finally
the other generators $t_{ij}$, $i\ne j$, in the usual order.
(This is a convenient choice, inessential for the outcome).
Using \rmat~, $t_{11} t_{22} t_{33}$
can be expressed as a polynomial in $t_{ij}$ of degree at most one in any of
the generators $t_{11}$, $t_{22}$ and $t_{33}$.
Unfortunately, by reordering we may raise the degree in the diagonal
generators. However, by analysing the commutation relations \coma - \comca~
we see that some of them (of the exchange type) are innoquous.
Those which are of the commutator type and between the diagonal  generators
produce only the non-diagonal generators $t_{ij}$ and lower the degree of
{\sl two} diagonal  generators.
Next, those of the commutator type between the non-diagonal generators
produce at most one  diagonal generator (namely $t_{22}$).
Altogether, there is a net lowering of the complexive degree of the product
$t_{11} t_{22} t_{33}$.
By iteration, this allows to lower consistently this degree up to reaching
the situation when one of the diagonal generators is not present.
Finally, by excluding the overlapping cases we arrive at the following result:

\noindent
The independent basis in \sL consist of three sectors composed by the elements
$t^{\bar n}$ with one of the following restrictions:
\eqn\base{\eqalign{
&(I) 	\ \ \ \  	n_{11}=0 \cr
&(II) 	\ \ \		n_{11}\ge 1, n_{22}=0 \cr
&(III) 	\ \		n_{11}\ge 1, n_{22} \ge 1, n_{33}=0 \ .\cr} }
Now, by iterating the twisted derivation rules for \sL
\eqn\lder{ {\cal L} (l^{\pm}_{ij}) \varphi \psi  =
\sum_k {\cal L} (l^{\pm}_{kj}) \varphi \cdot {\cal L} (l^{\pm}_{ik}) \psi \ ,}
\eqn\rder{ {\cal R} (l^{\pm}_{ij})\varphi\psi  =
\sum_k {\cal R} (l^{\pm}_{ik}) \varphi \cdot {\cal R} (l^{\pm}_{kj}) \psi \ ,}
we obtain the action of the representation ${\cal L}$ on $t^{\bar n}$
which we give in the Appendix.
\newsec{Reduction and intertwiners.}
We impose the conditions of the (infinitesimal) right covariance on the
independent basis for ~\SL~.
First, we consider only those functions $\psi $
(i.e. the multilabels $\bar n$) on which
\eqn\rcov{ {\cal R} (l^{+}_{ij})\psi  = 0\ , \ i<j \ . }
By these conditions some of the indices $n_{ij}$ have to vanish,
but for some other there is a possibility of compensation between
various pieces, c.f. (A 10-12).
This requires solutions to be of the form of infinite series.
Instead, we shall assume invertibility of some
(special combinations of) generators $t_{ij}$ and may recover
the series by formal expansion of the inverses in $q$.
By a deeper look, we see that the solutions of \rcov~ have to be built from
generators $t_{i3}$ in the last column and from minors $m_{i1}$
of the first column
($m_{ij} $ is the minor of $t_{ij} $, i.e. the q-determinant of T with
$i^{th}$ line and $j^{th}$ column removed).
Due to \det~ one of them is dependent on the rest and we choose to delete
$m_{31}$.  Next, we impose that on $\psi $
\eqn\rcovc{ {\cal R} (l^{+}_{ii})\psi  = \exp r_i ~\psi  \ ,}
for $i=1, 2$ (the case $i=3$ is not independent due to \detl~), where
the labelling numbers $r_1$ and of $r_2$ will be specified further.
We note the multiplicative property of the eigenvalue of ${\cal R}
(l^{+}_{ii})$
with respect to the product of eigenvectors and note that
consequently the quotients $t_{i3} t_{j3}^{-1}$ and $m_{i1} m_{j1}^{-1}$
have eigenvalue 1.
Among them, we choose to work with the following three independent
combinations:
\eqn\defw{ w_1 = t_{13} t_{33}^{-1}\ ,\
	   w_2 = t_{23} t_{33}^{-1}\ ,\
	   w_3 = m_{21} m_{11}^{-1} \ ,}
where $ m_{21} = t_{12} t_{33} -q t_{13} t_{32}$ and
$ m_{11} = t_{22} t_{33} -q t_{23} t_{32}$.

Instead, in order to solve \rcovc~, we employ the combination
$m_{11}^{j_1}~ t_{33}^{j_2} $, with $j_1, j_2 $ to be determined in terms of
$r_1, r_2$.  Indeed, we get
$$r_1 = -(j_2 + 2j_1)h/3\ , \ r_2 = -(j_1 + j_2)h/3\ , $$
where $q = \exp h$.
Thus, since $j_1, j_2 \in \bbz $,  the admissible values of $r_1$ and of $r_2$
have to be quantized in units of $h/3$.
Finally, the basis in the space ${\cal T}_{\bar j}$,
$\bar j =: (j_1, j_2)$,
of common solutions of \rcov~ and \rcovc~ consist of the ordered monomials:
\eqn\basisz{
\f =:
w_{1}^{n_1} w_{2}^{n_2} w_{3}^{n_3} m_{11}^{j_1} t_{33}^{j_2}  \ ,}
where $n_1, n_2, n_3 \in \bbz_+$.
In this way we obtain an infinite family of reduced representation spaces
${\cal T}_{\bar j}$, indexed by the integers $j_1, j_2$.

It is remarkable, that the three independent variables $w_i$
form a closed algebra:
\eqn\comrelw{
w_1 w_2 = q w_2 w_1 		,\
w_1 w_3 = q^{-1} w_3 w_1 	,\
w_2 w_3 = q w_3 w_2 + \l w_1 	.}
We give also some relations useful for the following computations:
\eqn\comrelww{
w_{3}^{n}  w_2 = q^{-n} w_2 w_{3}^{n} + \l q^{-1}[n] w_1 w_{3}^{n-1} ,\
w_3 w_{2}^{n}  = q^{-n} w_{2}^{n} w_3 + \l q^{-n}[n] w_1 w_{2}^{n-1} .}
Moreover, the two `spectators' $t_{33}$ and $m_{11}$ commute:
\eqn\comrels{
t_{33} m_{11} = m_{11} t_{33}~}
and have the following commutation relations with the variables $w_i$:
\eqn\comrelwm{
w_1 m_{11} = q m_{11} w_1 		 ,\
w_2 m_{11} = m_{11} w_2		 ,\
w_3 m_{11} = q m_{11} w_3 		 ,\ }
\eqn\comrelwt{
w_1 t_{33} = q t_{33} w_1		 ,\
w_2 t_{33} = q t_{33} w_2		 ,\
w_3 t_{33} = t_{33} w_3 		 .}
We note that our $w_i$ are also the relevant variables as far as the
global covariance is concerned. This can be seen from the Gauss decomposition
of $T$ :
\eqn\gauss{
 \pmatrix{ 	1 & w_{3} 		& w_{1} 		\cr
		0 & 1			& w_{2} 		\cr
		0 & 0			& 1			}\!
\pmatrix{ 	t_{11}\! -\! w_{3}t_{33}^{-1}m_{12}\! -\! w_{1}t_{31}&0&0\cr
		0&t_{33}^{-1} m_{11}&0					 \cr
		0&0&t_{33}						 }\!
\pmatrix{ 	1 			&0			&0\cr
		m_{11}^{-1} m_{12}	&1			&0 \cr
		t_{33}^{-1} t_{31} 	&t_{33}^{-1} t_{32}	&1}.}
By a lengthy but straightforward computation, iterating the twisted
derivation rules,
we obtain the explicit formulas for the representation in ${\cal T}_{\bar j}$~:
\eqn\lrepab{ {\cal L} (l^{+}_{12}) \f =
-\l q^{1-n_2-n_3+(2j_1+j_2)/3}[n_2] \phi^{n_1, n_2-1, n_3}_{\bar j}
+\l q^{1-n_2+(j_2-j_1)/3} [n_3\! -\! j_1] \phi^{n_1, n_2, n_3+1}_{\bar j} }
\eqn\lrepbc{ {\cal L} (l^{+}_{23}) \f =
\l q^{1-(2j_2+j_1)/3} [n_1\! +\! n_2\! -\! n_3\! -\! j_2]
\phi^{n_1, n_2+1, n_3}_{\bar j}
+\l q^{n_1-(2j_2+j_1)/3} [n_3] \phi^{n_1+1, n_2, n_3-1}_{\bar j} }
\eqn\lrepac{\eqalign{ {\cal L} (l^{+}_{13}) \f = &
\l q^{1-n_2+(4j_2-j_1)/3} [n_1\! -\! n_2\! -\! n_3\! -\! j_1\! +\! j_2]
\phi^{n_1+1, n_2, n_3}_{\bar j} \cr
&+\l q^{2-n_1-n_2+(j_2-j_1)/3} [j_1\! -\! n_3]
\phi^{n_1, n_2+1, n_3+1}_{\bar j} }}
\eqn\lrepaa{ {\cal L} (l^{+}_{11}) \f = q^{-n_1-n_3+(2j_1+j_2)/3} \f}
\eqn\lrepbb{ {\cal L} (l^{+}_{22}) \f = q^{-n_2+n_3+(j_2-j_1)/3} \f}
\eqn\lrepcc{ {\cal L} (l^{+}_{33}) \f = q^{n_1+n_2-(j_1+2j_2)/3} \f}
\eqn\lrepba{ {\cal L} (l^{-}_{21}) \f =
\l q^{n_1-1+(j_1-j_2)/3} [n_3] \phi^{n_1, n_2, n_3-1}_{\bar j}
+\l q^{n_2-n_3+(j_1-j_2)/3} [n_1] \phi^{n_1-1, n_2+1, n_3}_{\bar j} }
\eqn\lrepcb{ {\cal L} (l^{-}_{32}) \f =
\l q^{-1+(j_1+2j_2)/3} [n_2] \phi^{n_1, n_2-1, n_3}_{\bar j} }
\eqn\lrepca{ {\cal L} (l^{-}_{31}) \f =
\l q^{n_2-1+(j_1+2j_2)/3} [n_1] \phi^{n_1-1, n_2, n_3}_{\bar j} \ .}
So far we have restricted the left regular representation to (infinite
dimensional) subrepresentations ${\cal T}_{\bar j}$.
The restricted functions depend effectively only on the variables $w_i$
since for each fixed $\bar j$ the factor $m_{11}^{j_1}~t_{33}^{j_2}$ is fixed.
In order to pursue  the  reduction till the end and get the functions
depending {\sl manifestly} only on $w_i$ we associate with each
$\phi \in T_j$  a function $\hat \phi $, by the formula
\eqn\xhatfun{\hat\phi ~=~ \phi ~m_{11}^{-j_1}~t_{33}^{-j_2} \ .}
We note that we have actually a freedom to work with variables proportional
to $w_i$,
\eqn\xvar{z_i = C_i w_i,}
where $C_i$ are constants depending on $q$ and $\bar j$. (This leads to an
equivalent representation). In order to simplify the formulae we choose
\eqn\xvarc{C_1 = C_2 = q^{1-(j_1+2j_{2})/3},\ C_3 = q^{1+(j_2-j_1)/3}\ .}
The representations \lrepab - \lrepca~ acting in ${\cal T}_{\bar j}$,
induce the following transformation rules on the basis $\z$~:
\eqn\lrehab{ \hat{\cal L} (l^{+}_{12})  \z =
-\l q^{2-n_2-n_3+(j_1-j_2)/3} [n_2] z_{1}^{n_1} z_{2}^{n_2-1} z_{3}^{n_3 }
+ \l q^{-n_2} [n_3\! -\! j_1] z_{1}^{n_1 } z_{2}^{n_2} z_{3}^{n_3+1}}
\eqn\lrehbc{ \hat{\cal L} (l^{+}_{23})  \z =
\l q^{n_1+(j_2-j_1)/3} [n_3] z_{1}^{n_1+1 } z_{2}^{n_2} z_{3}^{n_3-1}
+ \l [n_1\! +\! n_2\! -\! n_3\! -\! j_2] z_{1}^{n_1 } z_{2}^{n_2+1 }
z_{3}^{n_3 }}
\eqn\lrehac{\eqalign{ \hat{\cal L} (l^{+}_{13})  \z = &
\l q^{-n_1-n_2+(2j_2+j_1)/3} [j_1\! -\! n_3] z_{1}^{n_1} z_{2}^{n_2+1}
z_{3}^{n_3+1} \cr
&+ \l q^{-n_2+2j_2} [n_1\! -\! n_2\! -\! n_3\! -\! j_1\! +\! j_2]
z_{1}^{n_1+1} z_{2}^{n_2}z_{3}^{n_3}}}
\eqn\lrehaa{ \hat{\cal L} (l^{+}_{11})  \z = q^{-n_1-n_3+(2j_1+j_2)/3} \z}
\eqn\lrehbb{ \hat{\cal L} (l^{+}_{22})  \z = q^{-n_2+n_3+(j_2-j_1)/3} \z}
\eqn\lrehcc{ \hat{\cal L} (l^{+}_{33})  \z = q^{n_1+n_2-(j_1+2j_2)/3} \z}
\eqn\lrehba{ \hat{\cal L} (l^{-}_{21})  \z =
\l q^{n_1} [n_3] z_{1}^{n_1} z_{2}^{n_2} z_{3}^{n_3 -1}
+ \l q^{n_2+n_3+(j_1-j_2)/3} [n_1] z_{1}^{n_1 -1} z_{2}^{n_2} z_{3}^{n_3}}
\eqn\lrehcb{ \hat{\cal L} (l^{-}_{32})  \z =
\l [n_2] z_{1}^{n_1} z_{2}^{n_2 -1} z_{3}^{n_3}}
\eqn\lrehca{ \hat{\cal L} (l^{-}_{31})  \z =
\l q^{n_2} [n_1] z_{1}^{n_1 -1} z_{2}^{n_2} z_{3}^{n_3}}
These expressions define the representations $\hat {\cal L}_j$ which act in the
(same for all $\bar j$) space $\hat {\cal T}_{\bar j}$ of polynomials
$\hat\phi$ in $z_i$ (which as a linear space is just $T_{00}$).
(Note however that the concrete form of $z_i$ depends on $\bar j$; we do not
indicate this dependence explicitly in order to simplify the notation).
Using the operations $M_i$, rescaling the variable $z_i\to q z_i$,
the q-derivative $D_{i} f = (M_i - M_{i}^{-1}) f/(q-q^{-1})z $
and the operator $Z_i$, multiplying by the variable $z_i$,
all of which by convention act
{\sl directly} on the $i^{th}$ place in the ordered monomials,
we obtain the following reduced q-differential representation~:
\eqn\lrezab{ \hat{\cal L} (l^{+}_{12}) =
-\l q^{2+(j_1-j_2)/3} M_{2}^{-1} M_{3}^{-1} D_{2} +
\l (q^{-j_1} M_{3} - q^{j_1} M_{3}^{-1}) M_{2}^{-1} Z_3 }
\eqn\lrezbc{ \hat{\cal L} (l^{+}_{23}) =
-\l q^{(j_2-j_1)/3} M_{1} Z_1 D_{3} +
\l (q^{-j_2} M_{1}^{-1}M_{2}^{-1}M_{3} - q^{j_2} M_{1}M_{2}M_{3}^{-1})Z_2 }
\eqn\lrezac{\eqalign{ \hat{\cal L} (l^{+}_{13}) = &
\l q^{(2j_2+j_1)/3} (q^{j_1} M_{3}^{-1} -
q^{j_1} M_{3})M_{1}^{-1}M_{2}^{-1} Z_2 Z_3\cr
& + \l q^{2j_2} (q^{j_2-j_1} M_{1}M_{2}^{-1}M_{3}^{-1} -
q^{j_1-j_2} M_{1}^{-1} M_{2} M_{3}) M_{2}^{-1} Z_1 } }
\eqn\lrezaa{ \hat{\cal L} (l^{+}_{11}) = q^{(2j_1+j_2)/3} M_{1}^{-1}M_{3}^{-1}}
\eqn\lrezbb{ \hat{\cal L} (l^{+}_{22}) = q^{(j_2-j_1)/3} M_{2}^{-1} M_{3}}
\eqn\lrezcc{ \hat{\cal L} (l^{+}_{33}) = q^{-(j_1+2j_2)/3} M_{1}M_{2}}
\eqn\lrezba{ \hat{\cal L} (l^{-}_{21}) =
\l M_1 D_{3} + \l q^{(j_1-j_2)/3} M_2 M_{3}^{-1} D_{1}}
\eqn\lrezcb{ \hat{\cal L} (l^{-}_{32}) = \l D_{2}}
\eqn\lrezca{ \hat{\cal L} (l^{-}_{31}) = \l M_2 D_{1}}
We remark, that although originally we have assumed $j_1, j_2 \in \bbz$,
the formulae \lrezab - \lrezca~ define a representation
$\hat {\cal T}_{\bar j}$ of \sL~ also for arbitrary $j_1, j_2 \in \bbc$.
However, only for $j_1, j_2 \in \bbz$ they comprise a deformation of
representations which are integrable to a representation of the group
$SL(3,\bbc)$.

Now we consider the right regular representation (A 16-18)
of the remaining generators on $\f$ and restrict it. It turns out that for any
$j_1, j_2$ there are always unwanted terms like $m_{12}$ or $m_{22}$
in the image of ${\cal R} (l^{-}_{31}) $
and we do not consider this operator any more.
Instead, we have
\eqn\rrepba{\eqalign{ {\cal R} (l^{-}_{21}) \f =
\l q^{(2j_1-j_1)/3}
(& q^{n_3-1} [n_3] w_{1}^{n_1}w_{2}^{n_2}w_{3}^{n_3-1}
m_{11}^{j_1-2}t_{33}^{j_2+1}\cr
&- [j_1] w_{1}^{n_1}w_{2}^{n_2}w_{3}^{n_3}
m_{11}^{j_1-1}t_{33}^{j_2}m_{12})
\ ,\cr}}
\eqn\rrepcb{\eqalign{ {\cal R} (l^{-}_{32}) \f =
-\l q^{(j_2-j_1)/3}
(& q^{n_1+n_2-n_3-1} [n_1] w_{1}^{n_1-1}w_{2}^{n_2}w_{3}^{n_3+1}
m_{11}^{j_1+1}t_{33}^{j_2-2}\cr
&+ q^{2n_1+n_2-n_3-1} [n_2] w_{1}^{n_1}w_{2}^{n_2-1}w_{3}^{n_3}
m_{11}^{j_1+1}t_{33}^{j_2-2}\cr
& + [j_2] w_{1}^{n_1}w_{2}^{n_2}w_{3}^{n_3}m_{11}^{j_1}t_{33}^{j_2-1}t_{32})
\ .\cr}}
Thus for ${\cal R} (l^{-}_{21}) $ and for ${\cal R} (l^{-}_{32}) $
the unwanted terms (with $t_{32}$ or $m_{12}$)
are not present precisely when $j_1=0$ and $j_2=0$, respectively.
Then, in fact, we obtain two intertwiners:
the restriction of ${\cal R} (l^{-}_{21}) $ to ${\cal T}_{0, j_2}$ which
maps into ${\cal T}_{-2, j_2+1}$ and the restriction of ${\cal R} (l^{-}_{32})$
to ${\cal T}_{j_1, 0}$ which maps into ${\cal T}_{j_1+1, -2}$.
By similar computations (first for the squares of these operators and then by
induction) one obtains a family of other intertwiners
\eqn\intba{ {\cal R} (l^{-}_{21})^{j_1+1} : {\cal T}_{j_1, j_2}
\to {\cal T}_{-j_1-2, j_2+j_1+1},~ {\rm for} ~~j_1\in\bbz_+~,~j_2\in\bbc  }
\eqn\intcb{ {\cal R} (l^{-}_{32})^{j_2 +1} : {\cal T}_{j_1, j_2 }
\to {\cal T}_{j_1+j_2 +1, -j_2 -2},~ {\rm for} ~~j_2 \in\bbz_+~,~j_1\in\bbc }
There are also some mixed intertwiners~:
\eqn\intcbba{ {\cal R} (l^{-}_{32})^{j_2 +1} {\cal R} (l^{-}_{21})^{j_1+1}
: {\cal T}_{j_1, j_2 -j_1-1} \to {\cal T}_{j_2 -j_1-1, -j_2 -2},~
{\rm for} ~~j_1, j_2 \in\bbz_+ ,}
\eqn\intbacb{ {\cal R} (l^{-}_{21})^{j_1+1} {\cal R} (l^{-}_{32})^{j_2 +1}
: {\cal T}_{j_1-j_2 -1, j_2 } \to {\cal T}_{-j_1-2, j_1-j_2 -1},~
{\rm for}~~j_1, j_2 \in\bbz_+ .}
In the same way as for the left-represented operators
we perform the explicit reduction to the space $\hat {\cal T}_{\bar j}$
of functions of the variables $z_i$. Keeping track of the changes of indices
$\bar n$ and of the labels $\bar j$, on which the coefficients $C_i$ depend,
we obtain the two basic intertwiners
\eqn\zintba{ \hat {\cal R} (l^{-}_{21}) \z = \l q^{1+2j_{2}/3} D_{3} \ ,}
\eqn\zintcb{ \hat {\cal R} (l^{-}_{32}) \z =
-\l q^{1-j_1/3} D_{1} Z_3	- \l q^{1-2j_1/3} M_1 D_{2} \ .}
The other intertwiners can be easily obtained by observing that the explicit
formula for the intertwiner corresponding to product of (any number of)
$l^{-}_{21}$ and $l^{-}_{32}$ is just the product of the basic intertwiners
\eqn\zint{ \hat {\cal R} \bigl( (l^{-}_{32})^{j_2}  (l^{-}_{21})^{j_1} \bigr) =
\bigl( \hat {\cal R} (l^{-}_{32})\bigr)^{j_2 }
\bigl( \hat {\cal R} (l^{-}_{21})\bigr)^{j_1}\ .}
They clearly satisfy
\eqn\interba{
\bigl( \hat {\cal R} (l^{-}_{21})\bigr)^{j_1+1}\
\cdot
\hat {\cal L}_{j_1, j_2}(l^{\pm}_{k\ell })
=
\hat {\cal L}_{-j_1-2, j_2+j_1+1}(l^{\pm}_{k\ell })
\cdot
\bigl( \hat {\cal R} (l^{-}_{21})\bigr)^{j_1+1}\ ,}
\eqn\intercb{
\bigl( \hat {\cal R} (l^{-}_{32})\bigr)^{j_2 +1}\
\cdot
\hat {\cal L}_{j_1, j_2 }(l^{\pm}_{k\ell })
=
\hat {\cal L}_{j_1+j_2 +1, -j_2 -2}(l^{\pm}_{k\ell })
\cdot
\bigl( \hat {\cal R} (l^{-}_{32})\bigr)^{j_2 +1}\ ,}
\eqn\interbacb{
\hat {\cal R} \bigl( (l^{-}_{21})^{j_1+1} (l^{-}_{32})^{j_2 +1} \bigr)
\cdot
\hat {\cal L}_{j_1-j_2 -1, j_2 }(l^{\pm}_{k\ell })
=
\hat {\cal L}_{-j_1-2, j_1-j_2 -1}(l^{\pm}_{k\ell })
\cdot
\hat {\cal R} \bigl( (l^{-}_{21})^{j_1+1} (l^{-}_{32})^{j_2 +1} \bigr) \ ,}
\eqn\intercbba{
\hat {\cal R} \bigl( (l^{-}_{32})^{j_2 +1} (l^{-}_{21})^{j_1+1} \bigr)
\cdot
\hat {\cal L}_{j_1, j_2 -j_1-1}(l^{\pm}_{k\ell })
=
\hat {\cal L}_{j_2 -j_1-1, -j_2 -2}(l^{\pm}_{k\ell })
\cdot
\hat {\cal R} \bigl( (l^{-}_{32})^{j_2 +1} (l^{-}_{21})^{j_1+1} \bigr)\ ,}
where we have explicitly indicated the labels of the representations
$\hat {\cal L}$.
These intertwiners give rise to a various partial equivalences between
some $\hat {\cal T}_{\bar j}$. In particular, the kernels of these intertwiners
form invariant subrepresentations.
Thus, for generic $j_1, j_2 \in\bbc$ the representations
$\hat {\cal T}_{\bar j}$ are irreducible except for $j_1\in \bbz_+$ or
$j_1\in \bbz_+$, when they are reducible.
To study further the reducibility, we note that all the intertwiners which
act on $\hat {\cal T}_{\bar j}$ for given $j_{1}, j_{2} \in \bbz_+$,
contain as a last factor either the $j_1+1$ power of the basic intertwiner
\zintba~ or the $j_2+1$ power of \zintcb~.
Thus it is sufficient to consider only the kernels of two intertwiners
\interba~ and \intercb~.
It turns out that their common intersection are irreducible
(sub)representations and have the dimension
$d=(j_1\! +\! 1) (j_2\! +\! 1) \bigl( 1\! +\! (j_1\! +\! j_2)/2\bigr) $.
They consist of polynomials in $z_i$ of limited order, depending on $i$ and on
$\bar j$, and are q-deformation of the well known representations of $sl(3)$.
In particular, $\{ 1, ~z_1, ~z_2 \}$ is a basis of
$3_q = {\rm ker}\hat {\cal R} (l^{-}_{21}) \cap
{\rm ker} \bigl( \hat {\cal R} (l^{-}_{32})\bigr)^2
\subset \hat {\cal T}_{0, 1} $,
$\{ 1, ~z_3, ~z_2z_3\! -\! q^{-1/3}z_1 \}$ is a basis of
$3_{q}^{*} = {\rm ker} \bigl( \hat {\cal R} (l^{-}_{21})\bigr)^2
\cap {\rm ker}\hat {\cal R} (l^{-}_{32})
\subset \hat {\cal T}_{1, 0} $
and $\{ 1, ~z_1, ~z_2, ~z_3, ~z_1z_3, ~z_2z_3,
{}~q^{1/3}(1\! +\! q)z_{2}^{2}z_3-[2]z_1z_2,
{}~q^{1/3}[2]z_1z_2z_3 \!-\! (1\! +\! q)z_{1}^{2} \}$ is a basis of
$8_{q} = {\rm ker} \bigl( \hat {\cal R} (l^{-}_{21}) \bigr)^2
\cap {\rm ker} \bigl( \hat {\cal R} (l^{-}_{32}) \bigr)^2
\subset \hat {\cal T}_{1, 1} $.
\newsec{Final remarks.}
In this paper, the method previously applied to simpler examples,
has been extended to \SL~. It permits to obtain in a canonical way some
important (reducible and irreducible) representations of \sL~ and their
intertwiners, and links intimately to what should be a q-analogue of the
(infinitesimal)  Borel-Weil construction.
The reduction procedure corresponds in fact to working on the quantum three
dimensional complex flag manifold $F_q(1,2;3)$ (the quotient manifold of
\SL \ by the Borel subgroup), c.f. \ref\S{Y. Soibelman, RIMS-780, (1991)},
\ref\TT{E. Taft, J. Towber, J. Algebra {\bf 142} (1991) 1.}.
This is indeed the case for the representations inside the Weyl chamber
($j_1\cdot j_2 \ne 0$), while the representations with
$j_1$ or $j_2 = 0$ live, as in the previous simpler examples,
over the q-projective spaces (or q-grassmannians).
Our functions are local representatives of global sections, like e.g.
$m_{11}^{j_1}~t_{33}^{j_2}$, of some quantum line bundles over $F_q(1,2;3)$.
This holds over one particular patch, coordinatized by the variables $z_i$
(or $w_i$), corresponding to the requirement of invertibility of $m_{11}$
and $t_{33}$.

We mention that this approach which relies on the infinitesimal covariance
and the use of representations of $g_q$, though closely related,
seems to be conceptually simpler than what should be a q-analogue of induced
representations (i.e. working with the global covariance under the
co-representations of $G_q$ ), see for example
\ref\BM{T. Brzezi\' nski, S. Majid,
Commun. Math. Phys. {\bf 157} (1993) 591.},
\ref\BK{R. Budzy\' nski, W. Kondracki, Quantum principal fibre bundles:
topological aspects, Warsaw preprint IM PAN 517, 1993.},
\ref\D{M. Durdevi\' c, Quantum principal bundles, Belgrade Univ. Preprint,
1993 (hep-th/9311029).},
\ref\P{P. Podle\' s, Symmetries of quantum spaces. Subgroups and quotient
spaces of quantum $SU(2)$ and $SO(3)$ groups, Warsaw Univ. preprint 1994.}.
A detailed comparison of these two methods and the geometry of the quantum
homogeneous space $F_q(1,2;3)$ will be discussed in the forthcoming paper.
\bigskip\bigskip\noindent
{\bf Acknowledgments}
\medskip
We thank Prof. C. Reina for helpful discussions.
\bigskip\noindent
{\bf Note added}
The Heisenberg realization of $U_qsl(n)$ in terms of q-difference operators
on the flag manifold is presented in
\ref\ANO{H. Awata, M. Noumi, S. Odake, Lett. Math. Phys. {\bf 30} (1994) 35.}.
After submitting this paper, we have been also informed about
\ref\Dobr{V.K. Dobrev, ASI-TPA/10/93 preprint (hep-th/9405150)},
which has overlap with our results.

\vfill
\eject

{\bf Appendix.}

Denote $s = {1\over 3}(n_{11}+n_{12}+\dots +n_{33})$.
The action of the representation ${\cal L}$ on
$t^{\bar n} = t_{11}^{n_{11}}~ t_{12}^{n_{12}} \dots t_{33}^{n_{33}} $
is given as follows:
$$
\eqalign{ {\cal L} (l^{+}_{12}) \t = &
-\l q^s\bigl(
q^{1-n_{12}-n_{13}-n_{21}} [n_{21}]
t_{11}^{n_{11}+1} \dots t_{21}^{n_{21}-1} \dots  t_{33}^{n_{33}}\cr
&+ q^{1-n_{21}-n_{22}-n_{13}} [n_{22}]
t_{11}^{n_{11}}t_{12}^{n_{12}+1}\dots t_{22}^{n_{22}-1}\dots t_{33}^{n_{33}}\cr
&+ q^{1-n_{21}-n_{22}-n_{23}} [n_{23}]
t_{11}^{n_{11}}\dots t_{13}^{n_{13}+1}\dots t_{23}^{n_{23}-1}\dots
t_{33}^{n_{33}}
\bigr)\cr}\eqno(A1)$$
$$\eqalign{ {\cal L} (l^{+}_{23}) \t =&
-\l q^s \bigl(
q^{1-n_{22}-n_{23}-n_{31}} [n_{21}]
t_{11}^{n_{11}} \dots t_{21}^{n_{21}+1} \dots t_{31}^{n_{31}-1} \dots
t_{33}^{n_{33}}\cr
&+ q^{1-n_{23}-n_{31}-n_{32}} [n_{32}]
t_{11}^{n_{11}}\dots t_{22}^{n_{22}+1}\dots t_{32}^{n_{32}-1}
t_{33}^{n_{33}}\cr
&+ q^{1-n_{21}-n_{32}-n_{33}} [n_{33}]
t_{11}^{n_{11}}\dots t_{23}^{n_{23}+1}\dots t_{33}^{n_{33}-1}
\bigr)\cr}\eqno(A2)$$
$$
\eqalign{ {\cal L} (l^{+}_{13}) \t & =
-\l q^{1-n_{31}+s} \bigl(
q^{-n_{12}-n_{13}-n_{21}} [n_{31}]
t_{11}^{n_{11}+1} \dots t_{31}^{n_{31}-1} \dots t_{33}^{n_{33}}\cr
& + q^{-n_{13}-n_{22}-n_{32}} [n_{32}]
t_{11}^{n_{11}} t_{12}^{n_{12}+1}\dots t_{32}^{n_{32}-1} t_{33}^{n_{33}}\cr
& + q^{-n_{23}-n_{32}-n_{33}} [n_{33}]
t_{11}^{n_{11}}\dots t_{13}^{n_{13}+1}\dots t_{33}^{n_{33}-1}
\bigr)\cr
& + \l^2 q^{1-n_{31}+s}\bigl(
q^{1-n_{21}-n_{22}-n_{13} -n_{12} } [n_{31}] [n_{22}]
t_{11}^{n_{11}+1} \dots t_{21}^{n_{21}+1} t_{22}^{n_{22}-1} \dots
t_{31}^{n_{31}-1} \dots t_{33}^{n_{33}}\cr
&+ q^{1-n_{22}-n_{23}} [n_{31}][n_{23}]
t_{11}^{n_{11}} \dots t_{13}^{n_{13}+1} t_{21}^{n_{21}+1}\dots
t_{23}^{n_{23}-1} \dots t_{31}^{n_{31}-1}\dots t_{33}^{n_{33}}\cr
& + q^{1-n_{23}-n_{32}} [n_{23}] [n_{32}]
t_{11}^{n_{11}}\dots t_{13}^{n_{13}+1}\dots t_{22}^{n_{22}+1} t_{23}^{n_{23}-1}
\dots t_{32}^{n_{32}-1} t_{33}^{n_{33}}
\bigr)\cr}\eqno(A3)$$
$$
{\cal L} (l^{+}_{11}) \t =
q^{s-n_{11}-n_{12}-n_{13}} \t\eqno(A4)$$
$$
{\cal L} (l^{+}_{22}) \t =
q^{s-n_{21}-n_{22}-n_{23}} \t\eqno(A5)$$
$$
{\cal L} (l^{+}_{33}) \t =
q^{s-n_{31}-n_{32}-n_{33}} \t\eqno(A6)$$
$$\eqalign{ {\cal L} (l^{-}_{21}) \t = &
\l q^{-s}\bigl(
q^{n_{11}+n_{12}+n_{23}} [n_{13}]
t_{11}^{n_{11}} \dots t_{13}^{n_{13}-1} \dots  t_{23}^{n_{23}+1}
t_{33}^{n_{33}}\cr
& + q^{n_{11}+n_{22}+n_{23}} [n_{12}]
t_{11}^{n_{11}} t_{12}^{n_{12}-1}\dots t_{22}^{n_{22}+1}\dots
t_{33}^{n_{33}}\cr
& + q^{n_{21}+n_{22}+n_{23}} [n_{11}]
t_{11}^{n_{11}-1}\dots t_{21}^{n_{21}+1}\dots t_{33}^{n_{33}}
\bigr)\cr}\eqno(A7)$$
$$
\eqalign{ {\cal L} (l^{-}_{32}) \t =&
\l q^{-s}\bigl(
q^{n_{21}+n_{22}+n_{33}} [n_{23}]
t_{11}^{n_{11}} \dots t_{23}^{n_{23}-1} \dots  t_{33}^{n_{33}+1}\cr
+& q^{n_{21}+n_{32}+n_{33}} [n_{22}]
t_{11}^{n_{11}} \dots t_{22}^{n_{22}-1}\dots
t_{32}^{n_{32}+1}t_{33}^{n_{33}}\cr
+& q^{n_{31}+n_{32}+n_{33}} [n_{21}]
t_{11}^{n_{11}}\dots t_{21}^{n_{21}-1}\dots t_{31}^{n_{31}+1}\dots
t_{33}^{n_{33}}
\bigr)\cr} \eqno(A8)$$
$$
\eqalign{ {\cal L} (l^{-}_{31}) \t & =
\l q^{n_{33}-s}\bigl(
q^{n_{11}+n_{12}+n_{23}} [n_{13}]
t_{11}^{n_{11}} \dots t_{13}^{n_{13}-1} \dots t_{33}^{n_{33}+1}\cr
& + q^{n_{11}+n_{22}+n_{32}} [n_{12}]
t_{11}^{n_{11}} t_{12}^{n_{12}-1}\dots t_{32}^{n_{32}+1} t_{33}^{n_{33}}\cr
&+ q^{n_{21}+n_{31}-n_{32}} [n_{11}]
t_{11}^{n_{11}-1}\dots t_{31}^{n_{31}+1}\dots t_{33}^{n_{33}}
\bigr)\cr
& + \l^2 q^{n_{33}-s}\bigl(
q^{n_{11}+n_{22}} [n_{12}] [n_{23}]
t_{11}^{n_{11}} \dots t_{12}^{n_{12}-1}\dots t_{22}^{n_{22}+1}
t_{23}^{n_{23}-1}
\dots t_{33}^{n_{33}+1}\cr
& + q^{n_{21}+n_{22}} [n_{11}][n_{23}]
t_{11}^{n_{11}-1} \dots t_{21}^{n_{21}+1} \dots t_{23}^{n_{23}-1}\dots
t_{33}^{n_{33}+1}\cr
& + q^{n_{21}+n_{32}} [n_{11}] [n_{22}]
t_{11}^{n_{11}-1}\dots t_{21}^{n_{23}+1} t_{22}^{n_{22}-1}\dots
t_{32}^{n_{32}+1} t_{33}^{n_{33}}
\bigr)\cr}\eqno(A9)$$
\bigskip
$$\eqalign{ {\cal R} (l^{+}_{12}) \t = &
\l q^{n_{32}-s}\bigl(
q^{n_{11}+n_{21}} [n_{31}]
t_{11}^{n_{11}} \dots t_{31}^{n_{31}-1} t_{32}^{n_{32}+1 }t_{33}^{n_{33}}\cr
& + q^{n_{11}+n_{22}} [n_{21}]
t_{11}^{n_{11}}\dots t_{21}^{n_{21}-1}t_{22}^{n_{22}+1}\dots t_{33}^{n_{33}}\cr
& + q^{n_{12}+n_{22}} [n_{11}]
t_{11}^{n_{11}-1} t_{12}^{n_{12}+1}\dots t_{33}^{n_{33}}
\bigr)\cr}\eqno(A10)$$
$$
\eqalign{ {\cal R} (l^{+}_{23}) \t = &
\l q^{n_{33}-s}\bigl(
q^{n_{12}+n_{22}} [n_{32}]
t_{11}^{n_{11}} \dots t_{32}^{n_{32}-1} t_{33}^{n_{33}+1}\cr
& + q^{n_{12}+n_{23}} [n_{22}]
t_{11}^{n_{11}}\dots t_{22}^{n_{22}-1} t_{23}^{n_{23}+1}\dots
t_{33}^{n_{33}}\cr
& + q^{n_{13}+n_{23}} [n_{12}]
t_{11}^{n_{11}} t_{12}^{n_{12}-1} t_{13}^{n_{13}+1} \dots t_{33}^{n_{33}}
\bigr)\cr}\eqno(A11)$$
$$
\eqalign{ {\cal R} (l^{+}_{13}) \t & =
\l q^{n_{33}-s}\bigl(
q^{n_{11}+n_{21}+n_{32}} [n_{31}]
t_{11}^{n_{11}} \dots t_{31}^{n_{31}-1} \dots t_{33}^{n_{33}+1}\cr
& + q^{n_{11}+n_{22}+n_{23}} [n_{21}]
t_{11}^{n_{11}} \dots t_{21}^{n_{21}-1}\dots t_{23}^{n_{23}+1} \dots
t_{33}^{n_{33}}\cr
& + q^{n_{13}+n_{12}+n_{23}} [n_{11}]
t_{11}^{n_{11}-1}\dots t_{13}^{n_{13}+1}\dots t_{33}^{n_{33}}
\bigr)\cr
& + \l^2 q^{n_{33}-s}\bigl(
q^{n_{11}+n_{22}} [n_{21}] [n_{32}]
t_{11}^{n_{11}} \dots t_{21}^{n_{21}-1} t_{22}^{n_{22}+1} \dots
t_{32}^{n_{32}-1} \dots t_{33}^{n_{33}+1}\cr
& + q^{n_{12}+n_{22}} [n_{11}] [n_{32}]
t_{11}^{n_{11}-1} t_{12}^{n_{12}+1} \dots t_{32}^{n_{32}-1}
t_{33}^{n_{33}+1}\cr
& + q^{n_{12}+n_{23}} [n_{11}] [n_{22}]
t_{11}^{n_{11}-1} t_{12}^{n_{12}+1}\dots t_{22}^{n_{22}-1} t_{23}^{n_{23}+1}
\dots t_{33}^{n_{33}}
\bigr)\cr}\eqno(A12)$$
$$
{\cal R} (l^{+}_{11}) \t =
q^{-s+n_{11}+n_{21}+n_{31}} \t\eqno(A13)$$
$$
{\cal R} (l^{+}_{22}) \t =
q^{s+n_{12}+n_{22}+n_{33}} \t\eqno(A14)$$
$$
{\cal R} (l^{+}_{33}) \t =
q^{s+n_{13}+n_{23}+n_{33}} \t\eqno(A15)$$
$$
\eqalign{ {\cal R} (l^{-}_{21}) \t = &
-\l q^{1-n_{12}+s}\bigl(
q^{-n_{21}-n_{31}} [n_{12}]
t_{11}^{n_{11}+1} t_{12}^{n_{12}-1} \dots  t_{33}^{n_{33}}\cr
& + q^{-n_{22}-n_{31}} [n_{22}]
t_{11}^{n_{11}} \dots t_{21}^{n_{21}+1} t_{22}^{n_{22}-1}\dots
t_{33}^{n_{33}}\cr
& + q^{-n_{22}-n_{32}} [n_{32}]
t_{11}^{n_{11}}\dots t_{31}^{n_{31}+1}\dots t_{32}^{n_{32}-1} t_{33}^{n_{33}}
\bigr)\cr}\eqno(A16)$$
$$
\eqalign{ {\cal R} (l^{-}_{32}) \t = &
-\l q^{1-n_{13}+s}\bigl(
q^{-n_{22}-n_{32}} [n_{13}]
t_{11}^{n_{11}} t_{12}^{n_{12}+1} t_{13}^{n_{13}-1} \dots t_{33}^{n_{33}}\cr
& + q^{-n_{23}-n_{32}} [n_{23}]
t_{11}^{n_{11}} \dots t_{22}^{n_{22}+1} t_{23}^{n_{23}-1}\dots
t_{33}^{n_{33}}\cr
& + q^{-n_{23}-n_{33}} [n_{33}]
t_{11}^{n_{11}}\dots t_{32}^{n_{32}+1} t_{33}^{n_{33}-1}
\bigr)\cr}\eqno(A17)$$
$$
\eqalign{ {\cal R} (l^{-}_{31}) \t = &
-\l q^{1-n_{13}+s}\bigl(
q^{-n_{12}-n_{21}-n_{31}} [n_{13}]
t_{11}^{n_{11}+1} \dots t_{13}^{n_{13}-1} \dots t_{33}^{n_{33}}\cr
& + q^{-n_{22}-n_{23}-n_{31}} [n_{23}]
t_{11}^{n_{11}} \dots t_{21}^{n_{21}+1}\dots t_{23}^{n_{23}-1}\dots
t_{33}^{n_{33}}\cr
& + q^{-n_{23}-n_{32}-n_{33}} [n_{33}]
t_{11}^{n_{11}}\dots t_{31}^{n_{31}+1}\dots t_{33}^{n_{33}-1}
\bigr)\cr
& + \l^2 q^{2-n_{13}+s}\bigl(
q^{-n_{22}-n_{31}} [n_{13}] [n_{22}]
t_{11}^{n_{11}} t_{12}^{n_{12}+1} t_{13}^{n_{13}-1} t_{21}^{n_{21}+1}
t_{22}^{n_{22}-1}\dots  t_{33}^{n_{33}}\cr
& + q^{-n_{22}-n_{32}} [n_{13}][n_{32}]
t_{11}^{n_{11}} t_{12}^{n_{12}+1} t_{13}^{n_{13}-1}\dots t_{31}^{n_{13}+1}
t_{32}^{n_{32}-1} t_{33}^{n_{33}}\cr
& + q^{-n_{23}-n_{32}} [n_{23}] [n_{32}]
t_{11}^{n_{11}}\dots t_{22}^{n_{22}+1} t_{23}^{n_{23}-1}
t_{31}^{n_{31}+1} t_{32}^{n_{32}-1} t_{33}^{n_{33}}
\bigr)\ .\cr}\eqno(A18)$$

\listrefs
\vfil\eject
\end